\documentclass{article}

\usepackage[preprint]{arxiv}

\usepackage[utf8]{inputenc} 
\usepackage[T1]{fontenc}    
\usepackage{hyperref}       
\usepackage{url}            
\usepackage{booktabs}       
\usepackage{amsfonts}       
\usepackage{nicefrac}       
\usepackage{microtype}      
\usepackage{xcolor}         
\usepackage{cleveref}       
\usepackage{graphicx}
\usepackage{doi}

\usepackage{threeparttable}

\usepackage{authblk}

\title{Machine Learning to Detect Anxiety Disorders from Error-Related Negativity and EEG Signals}

\date{\vspace{-5.5ex}\texttt{Ramya.Chandrasekar@student.curtin.edu.au, \{Rakibul.Hasan, Shreya.Ghosh, Tom.Gedeon, Zakir.Hossain1\}@curtin.edu.au}}

\author[1]{Ramya Chandrasekar}
\author[1,2]{Md Rakibul Hasan}
\author[1]{Shreya Ghosh}
\author[1,3]{Tom Gedeon}
\author[1]{Md Zakir Hossain}
\affil[1]{School of Electrical Engineering, Computing and Mathematical Sciences, Curtin University, Bentley, WA 6102, Australia}

\affil[2]{Department of Electrical and Electronic Engineering, BRAC University, Dhaka 1212, Bangladesh}

\affil[3]{Óbuda University, Budapest, Hungary}

\renewcommand{\headeright}{}
\renewcommand{\undertitle}{\vspace{-5ex}}
\renewcommand{\shorttitle}{Machine Learning to Detect Anxiety Disorders}

\hypersetup{
hidelinks
}

\begin{document}
\maketitle
\begin{abstract}
Anxiety is a common mental health condition characterised by excessive worry, fear and apprehension about everyday situations. Even with significant progress over the past few years, predicting anxiety from electroencephalographic (EEG) signals, specifically using error-related negativity (ERN), still remains challenging. Following the PRISMA protocol, this paper systematically reviews 54 research papers on using EEG and ERN markers for anxiety detection published in the last 10 years (2013 -- 2023). Our analysis highlights the wide usage of traditional machine learning, such as support vector machines and random forests, as well as deep learning models, such as convolutional neural networks and recurrent neural networks across different data types. Our analysis reveals that the development of a robust and generic anxiety prediction method still needs to address real-world challenges, such as task-specific setup, feature selection and computational modelling. We conclude this review by offering potential future direction for non-invasive, objective anxiety diagnostics, deployed across diverse populations and anxiety sub-types. 
\end{abstract}

\keywords{machine learning \and deep learning \and EEG, error-related negativity \and anxiety \and detection}

\section{Introduction}
Anxiety is endemic to every person, with an occurrence rate of approximately 20\% \citep{who2017}. Between 2020 and 2022, over one in six people (17.2\% or 3.4 million people) aged 16 to 85 years experienced an anxiety disorder \citep{abs}. Anxiety is caused by changes in the situation, nervousness and common symptoms, including sweating, trembling and excessive worrying, which affect a person’s daily life. Anxiety disorders encompass a range of conditions, such as generalised anxiety disorder (GAD), panic disorder (PD), social anxiety disorder (SAD), obsessive-compulsive disorder (OCD), various phobia-related disorders, physical pain related protective behaviour \citep{li2020lstm,li2021plaan} and depression \citep{ghosh2021depression}. Current clinical approaches for diagnosing these disorders often suffer from limitations in accuracy and objectivity, relying heavily on self-reports, patient histories and clinical observations. These methods can be subjective and may not capture the nuanced neural and behavioural patterns associated with anxiety, leading to potential misdiagnoses. Recent research has shown promising results in using machine learning techniques to detect anxiety through physiological analysis \citep{abd2023}, such as respiration, electrocardiogram (ECG), photoplethysmography (PPG), electrodermal response (EDA) and electroencephalography (EEG), to identify patterns associated with anxiety states \citep{abd2023}. 

Machine learning techniques are increasingly employed in mental health to understand complex patterns \citep{meyer2015enhanced}. Machine learning models can be analysed in various data types, including physiological signals, behavioural patterns and self-reported symptoms, to identify patterns indicative of anxiety disorders \citep{sanei2007introduction, crawford2020behavioural}. Some commonly used machine learning models are support vector machine (SVM), random forest (RF), logistic regression, convolutional neural networks (CNNs) and recurrent neural networks (RNNs). These models have shown promise in analysing complex data modalities, including EEG signals, and medical imaging, for the diagnosis and prediction of mental health disorders \citep{mughal2020systematic}. Numerous reviews exist across mental health domains, including depression, anxiety and stress. However, there is a scarcity of comprehensive review papers specifically focusing on classifying anxiety disorders using EEG and error-related negativity (ERN) markers with machine learning models. For instance, \citet{al2020review} reviewed EEG, event-related potential (ERP) and brain connectivity in SAD, while \citet{michael2021eeg} systematically reviewed ERN and CRN (correct-response negativity) related to attentional control in anxiety disorders. Additionally, \citet{de2021deep} explored deep learning methods applied to EEG data across various mental disorders. Despite these efforts, there remains a notable absence of a holistic review covering various anxiety disorders using EEG and ERN with machine learning models.

Our approach adheres to the standard methodology of conducting a systematic literature review. We first formulated search keywords and queried eight databases, including Web of Science, Neuroscience and IEEE Xplore, yielding 986 papers. These papers underwent rigorous screening against three exclusion criteria through thorough title-and-abstract and full-text evaluations. Ultimately, 54 papers were selected for comprehensive review in this study. Our analysis is structured in two different ways: (1) EEG using machine learning models, and (2) ERN using statistical analysis. Our major contributions include:
\begin{enumerate}
  \item We provide an overview of the tasks and subjects utilised for data collection to leverage machine learning in anxiety detection.
  \item We thoroughly review all EEG and ERN-based machine learning models employed in various anxiety disorder studies published between 2013 and 2023.
  \item We offer detailed insights and future research direction into detection of various anxiety disorders, including GAD, SAD, OCD and PD, specifically examining EEG and ERN markers.
\end{enumerate}

\begin{figure}[!t]
    \centering
    \includegraphics[width=0.6\linewidth]{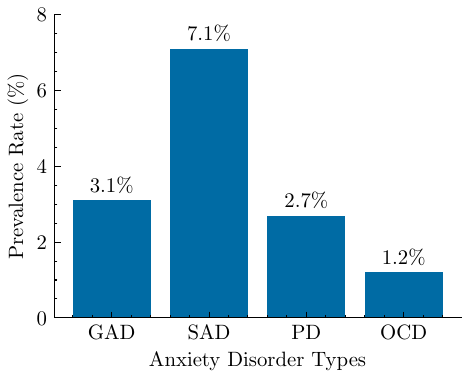}
    \caption{Annual prevalence rates of four major types of anxiety disorder \citep{nimh}. \textbf{Abbreviations:} GAD (Generalised anxiety disorder), SAD (Social anxiety disorder), OCD (Obsessive-compulsive disorder), PD (Panic disorder).}
    \label{fig:types}
\end{figure}

\section{Preliminaries}
\subsection{Types of Anxiety Disorders}
This paper focused on four major types of anxiety disorders: generalised anxiety disorder (GAD), social anxiety disorder (SAD), obsessive-compulsive disorder (OCD) and panic disorder (PD) \citep{nimh}. \autoref{fig:types} illustrates the worldwide annual prevalence rates of anxiety disorders. GAD arises from multiple sources, leading to pervasive fear and anxiety in affected individuals. This disorder causes sufferers to feel anxious even about everyday activities \citep{schacter2011psychology}. SAD is characterised by a fear of being judged in social situations. In these situations, individuals with social anxiety display more anxious behaviours and heightened autonomic arousal and report higher levels of distress compared to those without anxiety \citep{barker2015individual}. People with PD experience frequent and unexpected panic attacks. These attacks involve sudden surges of fear or discomfort, or a feeling of losing control, even when there is no obvious threat or cause \citep{nimh}. OCD is a chronic condition where a person has persistent and uncontrollable thoughts (obsessions) and engages in repetitive actions (compulsions), or both. These symptoms can be time-consuming, leading to significant distress and disruptions in daily life \citep{nimh}.

\subsection{Electroencephalogram (EEG)}
Electroencephalogram (EEG) is a non-invasive and cost-effective technique for measuring electrophysiological activity \citep{aldayel2024}. This non-invasive technique is well-suited for investigating the electrophysiological and cognitive conditions of the human brain \citep{aldayel2024}. It involves placing electrodes on the scalp to detect the neuronal activity. EEG has become a crucial tool for studying the dynamic patterns of brain activity and is increasingly used in clinical mental health assessments. Its potential extends to detecting various emotions, stress levels, anxiety and diverse brain disorders \citep{meyer2016}. EEG studies of anxiety disorders often focus on brain activity in the frontal lobe region. This area is associated with cognitive functions such as decision-making, emotion regulation and attentional control, which are often disrupted in individuals with anxiety disorders \citep{meyer2017biomarker, falkenstein1991effects}. Evaluating anxiety using EEG involves several steps: data collection, data pre-processing, feature extraction and detection of anxiety. Feature extraction and detection are the two main components of a standard EEG anxiety evaluation approach. There are three types of domain-based EEG features: time-domain, frequency-domain and time-frequency domain \citep{mazlan2024review}.

\subsection{Error-Related Negativity (ERN)}

Event-related potential or electrical brain response is time-locked to the specific event or stimuli obtained from EEG signals. It consists of various components that reflect different brain information stages \citep{gehring1993neural, brazdil2005intracerebral}. One specific ERP component is error-related negativity (ERN), which is a negative deflection in the EEG waveform that occurs within a specific window following the completion of the error. ERN is typically observed at frontocentral electrode sites and indicates the activity of the anterior cingulate cortex \citep{dehaene1994localization, hajcak2004error}. The negative peak in the EEG waveform occurs approximately (50--100 ms) after the commission of errors. Several studies have shown that the amplitude of the ERN is very sensitive to anxiety-related disorders \citep{hajcak2003anxiety, meyer2012development}. Larger ERN amplitude is associated with negative effects and transdiagnostic characteristics of anxiety disorders and is more pronounced in various anxiety disorders \citep{moser2013relationship, carrasco2012cognitive} such as OCD \citep{endrass2014performance, endrass2010performance}, SAD \citep{kujawa2016error, weinberg2012increased} and GAD \citep{xiao2011error, wiswede2009modulation}.

\section{Paper Screening Using PRISMA Method}

\subsection{Search Strategy}
To ensure the replicability of our study, we followed the PRISMA standard guidelines \citep{page2021prisma}. We conducted a computerised search strategy across multiple databases, including Google Scholar, ScienceDirect, IEEE Xplore, PubMed, ProQuest, Scopus, Neuroscience and Web of Science. The search terms covered all relevant keywords: ((“anxiety disorder” OR “GAD” OR “generalized anxiety disorder”) AND (“SAD” OR “Social anxiety disorder’) AND (“OCD” OR “Obsessive-compulsive disorder”) (“electroencephalogram” OR “EEG” OR “electroencephalographic”)) AND (“error-related negativity” OR “ERN” OR “event-related potential” OR “ERP”) AND ((“machine learning” OR “neural networks” OR “multilayer perceptron” OR “MLP’ OR “recurrent neural network” OR “RNN” OR “long short-term memory” OR “LSTM”) OR (“biomarkers” OR “physiological markers”)).
\begin{figure*}
    \centering
    \includegraphics[width=0.85\linewidth]{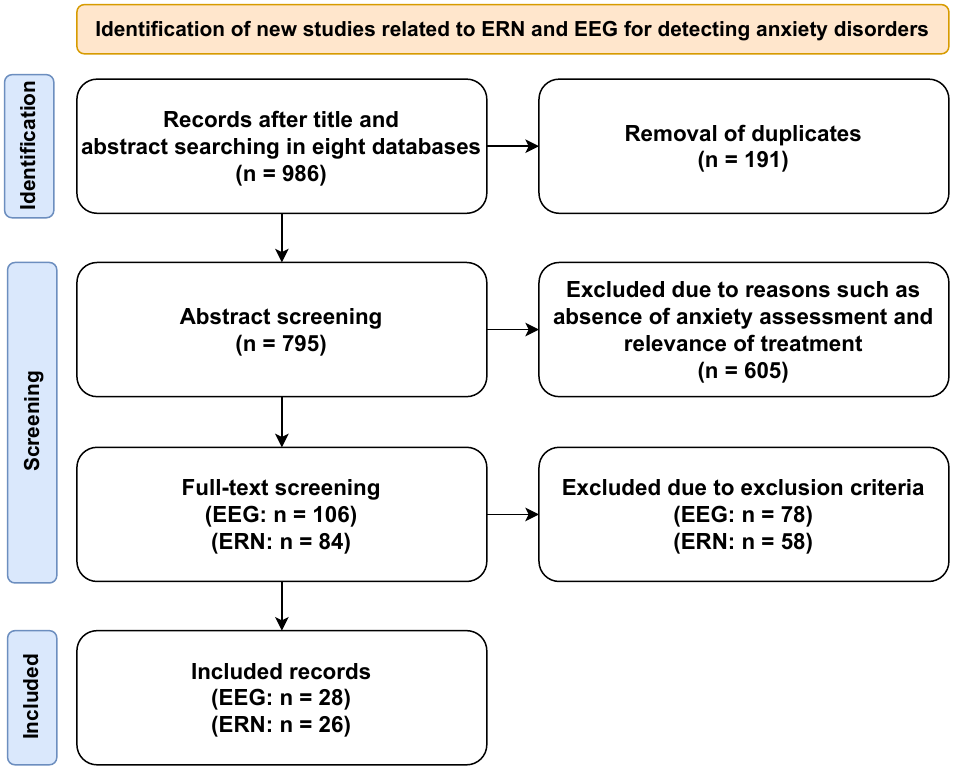}
    \caption{Flowchart for collecting and screening papers in our systematic review procedure based on PRISMA standard.}
    \label{fig:prisma}
\end{figure*}

\subsection{Inclusion/Exclusion Criteria}
Inclusion and exclusion criteria were established to ensure the selection of relevant studies for the systematic review. Studies were included if they met all three predefined criteria: (1) participants diagnosed with clinical anxiety disorders such as GAD, SAD, PD, OCD and any other anxiety disorders, (2) use of EEG and ERP measures related to anxiety disorders; and (3) studies conducted between 2013 and 2023. Furthermore, reviews summarising primary studies on mixed anxiety disorders, particularly those administered by clinicians, trained professionals or volunteer participants, were also included in the study selection process. Exclusion criteria were applied to filter out studies that did not meet the objectives of the review. Specifically, meta-analyses, systematic reviews, and review papers were excluded from the analysis. Additionally, studies addressing comorbidities of anxiety with depression and stress were excluded.

\subsection{Screening Result}
The search yielded 986 articles on anxiety across multiple databases, which included 191 duplicates. After abstract screening of 796 articles, 190 articles went to next stage for full-text screening. Among these, 106 articles focused on mixed anxiety disorder using EEG with machine learning models, and 183 articles were on mixed anxiety using various event-related potential components, including ERN. After the full-text screening phase, 28 met the inclusion criteria for EEG-based studies, while 26 met the criteria for ERP studies. \autoref{fig:prisma} illustrates the flow of number of papers in our screening process.


\section{Data Collection Strategies in Classifying Anxiety Disorders}
\autoref{tab:data} illustrates various data types that can be used to train machine learning algorithms, including information from questionnaires, interviews, demographic data, medical records, treatment histories and anxiety rating scales. This review focuses on wearable technology, demographics, channels, scale(s), and tasks that are used for the collection of EEG signals.

\begin{table*}[htbp]
\centerline{
\begin{threeparttable}
\scriptsize
\caption{Data collection of EEG and ERN}
\label{tab:data}
\begin{tabular}{@{}p{2cm}p{2.5cm}p{3cm}p{2.5cm}p{3cm}p{3cm}@{}}
\toprule
\textbf{Study} & \textbf{Population} & \textbf{Tasks} & \textbf{Channels} & \textbf{Wearables} & \textbf{Assessment Tools} \\ \midrule
\citep{gross2021machine} & Healthy control (HC) & Eye open and close & 62-channel & ActiCAP & STAI \\ 
\citep{minkowski2021feature} & AD vs HC & Awake & 66-channel & SynAmps2, NeuroScan & BDI-II, TAI \\ 
\citep{luo2024integrating} & GAD vs HC & Close, Awake, Relax & 16-channel & Nicolet EEG & DSM-V, HAM-A \\ 
\citep{mou2024prolonged} & AS vs GAD & Resting State & 16-channel & Nicolet EEG & HAM-A, DSM-V \\ 
\citep{wang2022difference} & Young GAD vs Old GAD & Close, Awake, Relax & 16-channel & Nicolet EEG & DSM-V, HAM-A \\ 
\citep{mohan2023classification} & Healthy Controls & Awake & 16-channel & BioSemi & SAM \\ 
\citep{shen2022aberrated} & GAD Patients & Resting State & 16-channel & Nicolet EEG & DSM-V, HAM-A \\ 
\citep{al2020review, al2022complexity} & GAD vs HC & Resting State & 16-channel & Nicolet EEG & DSM-V \\ 
\citep{al2023machine} & HC & Resting State & 32-channel & ANTNeuro & SIAS \\ 
\citep{arsalan2020electroencephalography} & HC & Awake State & Five-channels (Fz, FCz, Cz, FC1 and FC2) & Muse headband & STAI \\ 
\citep{muhammad2022human, aldayel2024, shikha2021stacked} & HC & Exposure Therapy (CBT) & 14-channel & Emotiv Epoc & HAM-A, SAM \\ 
\citep{weinberg2012increased} & GAD vs HC & Resting State & 16-channel & Nicolet EEG & HAM-A \\ 
\citep{yadawad2024predicting} & Healthy Controls & Relax, Speak, Question, Scary, Spell Bee & 14-channel & Emotiv Epoc & SAM \\ 
\citep{qi2023neuroimaging} & GAD vs DD & Resting State & 16-channel & Nicolet EEG & DSM-5, HAM-A \\
\citep{liu2023enhancing} & Healthy Patients & Eye open/close & 128-channel & Electrical Geodesic Instrument & Behavioral Data \\ 
\citep{aderinwale2023two} & HC vs PD & Rest, Stimulation, Recovery & Two-channel & Procomp Infiniti & HAM-A, SRI, PDSS \\ 
\citep{carrasco2012cognitive} & Pediatric OCD & Arrow flanker & Two electrode sites (FCz and Cz) & BioSemi ActiveTwo & CBCL, MASC, CDI \\ 
\citep{meyer2015enhanced} & Volunteer-child & Go/No-go, flanker, Stroop tasks & Five electrode sites (Fz, FCz, Cz, FC1 and FC2) & BioSemi ActiveTwo & SCARED \\ 
\citep{carrasco2013increased} & Pediatric anxiety & Arrow flanker & Two electrode sites (FCz and Cz) & BioSemi ActiveTwo & CBCL \\ 
\citep{kaczkurkin2013effect}& Volunteers & Letter flanker & 128-channel & Geodesics Sensor Cap & OCI-R \\ 
\citep{torpey2013error} & Volunteer-child & Go/No-go flanker & 32-channel & BioSemi ActiveTwo & DSM-IV \\ 
\citep{hum2013neural} & Clinically anxious children & Go/No-go flanker & 128-channel & Geodesics Sensor Cap & CBCL, MASC, STAIC-S \\ 
\citep{larson2013cognitive} & GAD/HC & Flanker Task & 128-channel & Geodesics Sensor Cap & BDI-II and STAI \\ 
\citep{rabinak2013neural} & Volunteers & Arrow flanker & 34-channel & BioSemi ActiveTwo & DSM-IV, CAPS, CES, BDI-II \\ 
\citep{zambrano2014differential} & Volunteers & Flanker Task & Two channels & Neuroscan Synamps & OCI-R, STAI-T, PSWQ \\ 
\citep{riesel2014overactive} & Volunteers & Go/No-go, flanker, Stroop tasks & 64-channel & BioSemi ActiveTwo & DSM-IV \\ 
\citep{weinberg2016error} & GAD/OCD/MDD/HC & Cognitive Task & 34-channel & BioSemi ActiveTwo  & DSM-IV, SCID, IMAS \\ 
\citep{kujawa2016error} & SAD/GAD/HC/AD & Flanker Task & 34-channel & BioSemi ActiveTwo  & DSM-IV \\ 
\citep{hanna2020diagnostic} & Clinical pediatric & Arrow flanker & 64-channel & BioSemi ActiveTwo  & CBCL \\ 
\citep{lo2017associations} & Volunteer-child & Go/No-go flanker & 64-channel & BioSemi ActiveTwo  & DSM-IV, RCADS-P \\ 
\citep{roh2017modulation} & Clinical-OCD & Arrow flanker & 64-channel & BioSemi ActiveTwo  & HAM-A, DOCSBDI, STAI \\ 
\citep{banica2019overprotective} & Volunteer-students & Flanker Task & 32-channel & BrainVision actiCHamp  & IDAS-II, STRAIN \\ 
\citep{cole2023relational} & Volunteer-child & Arrow flanker & 34-channel & BioSemi ActiveTwo  & MASC, SEQ \\ 
\bottomrule
\end{tabular}
\begin{tablenotes}
\small
\item \textbf{Abbreviations:} HC (Healthy Control), STAI (State-Trait Anxiety Inventory), AD (Alzheimer's Disease), BDI-II (Beck Depression Inventory-II), TAI (Test Anxiety Inventory), GAD (Generalized Anxiety Disorder), DSM-V (Diagnostic and Statistical Manual of Mental Disorders, Fifth Edition), HAM-A (Hamilton Anxiety Rating Scale), AS (Autism Spectrum), SAM (Self-Assessment Manikin), SIAS (Social Interaction Anxiety Scale), DD (Depression Disorder), DSM-5 (Diagnostic and Statistical Manual of Mental Disorders, Fifth Edition), Behavioral Data (Behavioral Data), PD (Parkinson's Disease), SRI (Social Readjustment Rating Scale), PDSS (Parkinson’s Disease Sleep Scale), CBCL (Child Behavior Checklist), MASC (Multidimensional Anxiety Scale for Children), CDI (Children's Depression Inventory), SCARED (Screen for Child Anxiety Related Disorders), OCI-R (Obsessive-Compulsive Inventory-Revised), DSM-IV (Diagnostic and Statistical Manual of Mental Disorders, Fourth Edition), CAPS (Clinician-Administered PTSD Scale), CES (Center for Epidemiologic Studies Depression Scale), PSWQ (Penn State Worry Questionnaire), SCID (Structured Clinical Interview for DSM Disorders), IMAS (Inventory of Multidimensional Anxiety Symptoms), SAD (Social Anxiety Disorder), RCADS-P (Revised Child Anxiety and Depression Scale - Parent Version), DOCSBDI (Dimensional Obsessive-Compulsive Scale, Beck Depression Inventory), IDAS-II (Inventory of Depression and Anxiety Symptoms-II), STRAIN (Social Threat, Ruminative Thoughts, and Anxiety Inventory), SEQ (Social Experiences Questionnaire).
\end{tablenotes}
\end{threeparttable}}
\end{table*}

\subsection{Electroencephalogram (EEG)}
EEG data were collected using a 16-channel system (Nicolet EEG TS215605) while subjects remained awake and relaxed with their eyes closed for ten minutes \citep{shen2022aberrated}. \citet{al2023machine} collected the data using a referential 32-channel cap (ANT Neuro) during a six-minute resting state with participants' eyes closed in a quiet, dimly-lit room. The channels covered prefrontal, temporal, parietal, and occipital regions. Participants were categorised based on SIAS scores for SAD and were all right-handed, healthy, and medication-free. \citet{muhammad2022human} utilised an Emotiv EPOC wireless headset with 14 electrodes during a 6-minute exposure therapy session involving anxiety-inducing scenarios and self-assessment tasks. Participants, 23 healthy adults (10 males, 13 females) with an average age of 30, reported anxiety levels using the HAM-A and SAM scales. The channels included AF3, F7, F3, FC5, T7, P7, O1, O2, P8, T8, FC6, F4, F8, and AF4 \citep{muhammad2022human}. 
 
\citet{mou2024prolonged} also used the HAM-A and DSM-V scales with a 16-electrode (Nicolet EEG TS215605) system but focused specifically on resting state tasks. Their research compared individuals with Anxiety disorders and those with GAD. \citet{wang2022difference} conducted another related study, using the DSM-V and HAM-A scales during close, awake, and relax tasks with the same (Nicolet EEG TS215605) system. They compared brain activity between younger and older GAD patients, contributing to an understanding of how age-related factors influence anxiety disorders \citep{wang2022difference}. Conversely, \citet{liu2023enhancing} took this a step further by employing a 128-channel Electrical Geodesic Instrument, although they focused on behavioural data related to eye-opening and closing tasks rather than anxiety-specific scales like DSM-V or HAM-A. This approach potentially allowed for more precise localisation of brain activity patterns, though it diverged from the anxiety-focused scales used in the other studies. Overall, these studies highlight the widespread use of the HAM-A and DSM-V scales in conjunction with EEG systems to explore the neural correlates of anxiety. The consistent use of resting state tasks, particularly with the Nicolet EEG system, underscores the importance of this approach in identifying biomarkers for anxiety disorders. Moreover, variations in EEG channel numbers across studies suggest that while more channels can offer greater spatial resolution, the 16-electrode setup remains a popular choice for balancing complexity with practical data collection and analysis.

\subsection{Error-Related Negativity (ERN)}
Several studies have employed the BioSemi ActiveTwo system to explore pediatric and volunteer populations' responses to various cognitive tasks. For example, \citet{carrasco2013increased} used this system with pediatric OCD and anxiety patients during an arrow flanker task, focusing on two electrode sites (FCz and Cz) and employing scales like the CBCL, MASC, and CDI to assess behavioural outcomes. Similarly, \citet{meyer2015enhanced} examined child volunteers using Go/No-go, flanker, and Stroop tasks with five electrode sites. Their study utilised the SCARED scale to measure anxiety and response control. \citet{riesel2014overactive}extended this methodology by using a 64-channel BioSemi system to investigate volunteers during Go/No-go and flanker tasks. They used the DSM-IV scale, indicating a broader exploration of anxiety-related responses across different tasks. Other studies have leveraged the Geodesics Sensor Cap with high channel counts to investigate anxiety-related responses. For example, \citet{kaczkurkin2013effect} used a 128-channel setup to study volunteers in a letter flanker task, focusing on Obsessive-Compulsive Inventory scores (OCI-R). Likewise, \citet{larson2013cognitive} examined GAD and healthy control populations during flanker tasks with a 128-channel system, employing both the BDI-II and STAI scales to assess anxiety and depression.

\citet{hum2013neural} also used a 128-channel Geodesics Sensor Cap with clinically anxious children during Go/No-go flanker tasks, combining CBCL, MASC and STAIC-S scales to measure anxiety severity and cognitive control.
The choice of systems and tasks varied significantly, with \citet{zambrano2014differential} using the two-channel Neuroscan Synamps system for flanker tasks, indicating a minimalist approach. In contrast, \citet{weinberg2012increased} employed a 34-channel BioSemi system to study GAD, OCD, and major depressive disorder (MDD) patients during cognitive tasks, broadening the scope of mental health research. Similarly, \citet{kujawa2016error} used a 34-channel BioSemi system for flanker tasks across several populations (SAD, GAD, HC, AD), focusing on the DSM-IV scale. Overall, these studies emphasise the importance of EEG systems like BioSemi and Geodesics Sensor Cap in exploring anxiety and cognitive control across various populations. The different scales (e.g., CBCL, STAI, DSM-IV) reflect a comprehensive approach to measuring psychological responses, providing valuable insights into the neural correlates of anxiety and related disorders.

\begin{table}[t]
\centering
\footnotesize
\caption{Summary of Feature Types and Corresponding Studies}
\label{tab:feature}
\begin{tabular}{@{}lp{7.5cm}@{}}
\toprule
\textbf{Feature Type}                        & \textbf{Studies}                                                                                         \\ \midrule
Power Spectral Density (PSD)                 & \citep{gross2021machine,shen2022aberrated, park2021identification, aldayel2024,muhammad2022human}         \\ 
Fuzzy Entropy                                & \citep{al2021analysis,shen2022aberrated}                                                                  \\ 
Time Domain Features                         & \citep{arsalan2020electroencephalography, fang2024exploring, shikha2021stacked, muhammad2022human}                          \\ 
Frequency Domain Features                    & \citep{muhammad2022human, shikha2021stacked}                                                                \\ 
Effective Connectivity (EC)                  & \citep{al2021analysis}                                                                                      \\ 
Phase Lag Index (PLI)                        & \citep{fang2024exploring}                                                                                        \\ 
Discrete Wavelet Transform (DWT)             & \citep{aldayel2024, baghdadi2019dasps, baghdadi2021psychological}                                                         \\ 
Fractal Dimension, Hjorth Parameters, HHS    & \citep{baghdadi2019dasps, baghdadi2021psychological}                                                                             \\ 
Lempel-Ziv Complexity, Correlation Dimension & \citep{aderinwale2023two}                                                                                  \\ 
Frontal Asymmetry Index (FAI)                & \citep{gross2021machine}                                                                                        \\ 
Functional Connectivity (FC)                 & \citep{park2021identification}                                                                                         \\ 
Recursive Feature Elimination                & \citep{shikha2021stacked, muhammad2022human}    
\\
\bottomrule
\end{tabular}
\end{table}

\section{Feature Extraction and Anxiety Detection}
\subsection{Error-Related Negativity (ERN)}
Feature selection was applied before feeding all features into the anxiety detection algorithms to explore whether there are subsets of features that offer improved detection. \citet{carrasco2013increased} examined enlarged ERN amplitudes as a neurophysiological feature in youth with OCD, GAD and SAD. ERN amplitude was compared between three groups using ANOVA with an error trail. \citet{meyer2015enhanced} assessed the ERN, CRN and $\Delta$ERN (difference between ERN and CRN) as neurophysiological features measured at electrodes Fz, Cz and Pz, alongside behavioural metrics like reaction times and error rates. Statistical analysis used repeated-measures ANOVAs and logistic regression to examine relationships between $\Delta$ERN, child anxiety disorders and maternal anxiety history. \citet{riesel2019error} analysed ERN and CRN amplitudes at electrodes Fz, Cz and Pz in OCD patients, OCD relatives and healthy controls, using ANCOVA and repeated-measures ANCOVA to account for age and response type. Hierarchical regressions explored the impact of family history on ERN in unaffected participants. Results showed enhanced ERN in OCD patients and relatives compared to control.

\citet{kujawa2016error} used mixed-design ANCOVAs to evaluate $\Delta$ERN pre-and post-treatment in anxiety patients (GAD and SAD) versus healthy controls, controlling for age and comorbid depression. Behavioural performance was assessed by accuracy and reaction time (RT), revealing slower RTs in patients compared to controls, though this difference was not significant post-treatment. \citet{lo2017associations} analysed EEG data from 64 electrodes to measure ERN and CRN, focusing on midline sites (Fz, FCz, Cz, CPz, Pz) with a 0--100 ms post-response window. Mixed-design ANOVAs revealed a significant ERN effect, with greater negativity for errors compared to correct responses, and a notable $\Delta$ERN at fronto-central sites. Behavioural measures showed faster responses on error trials and increased post-error accuracy. \citet{torpey2013error} assessed ERN, and Pe, which was measured across midline electrodes (Fz, Cz, Pz) and defined by average voltage during specific post-response windows (0--100 ms for ERN, 200--500 ms for Pe). The difference in error-related activity ($\Delta$ERN) was calculated by subtracting correct-trial from error-trial voltages. Simultaneous regression analyses assessed relationships between ERP components, parental psychopathology and child temperament, with age controlled for $\Delta$ERN. Results showed typical task performance, with faster reaction times for errors and maximal $\Delta$ERN at Cz. Overall, these key features analysed across studies include ERN, CRN and $\Delta$ERN amplitudes measured at midline electrodes (Fz, Cz, Pz), alongside behavioural metrics like reaction time and accuracy. Analytical methods involved mixed-design ANOVAs, ANCOVAs and regression models to assess relationships between ERN components, anxiety disorders, family history and treatment effects.

\subsection{Electroencephalogram (EEG)}
In recent years, several studies have explored different feature extraction techniques (\autoref{tab:feature}) in EEG-based machine learning models to improve the classification and detection of various neurological and psychological conditions. The features used for analysis span time, frequency, and connectivity domains, demonstrating a diverse approach to EEG signal processing.

One common feature extraction method is Power Spectral Density (PSD), which has been employed across multiple studies. For instance, \citet{gross2021machine} used PSD, frontal asymmetry index (FAI) and sub-band information to assess brain activity, while \citet{shen2022aberrated} combined PSD with univariate analysis, fuzzy entropy and multivariate functional connectivity to extract meaningful EEG features. Additionally, \citet{park2021identification} used PSD alongside functional connectivity (FC) at different frequency bands to study brain network alterations.

In addition to PSD, fuzzy entropy has emerged as a critical measure in EEG research. \citet{al2021analysis} utilised fuzzy entropy values as a primary feature extraction method, while \citet{shen2022aberrated} included fuzzy entropy in combination with functional connectivity measures. Fuzzy entropy helps quantify the complexity and unpredictability of EEG signals, which can be indicative of underlying neurological conditions. Time domain features have also been widely adopted, offering insight into the amplitude and signal characteristics over time. \citet{arsalan2020electroencephalography}and \citet{fang2024exploring} leveraged time-domain features, while \citet{muhammad2022human} focused on extracting frequency domain features such as mean power, rational asymmetry and asymmetry index. By concentrating on the theta and beta bands, they were able to pinpoint critical frequency components using a frequency band selection algorithm.

Other studies have explored more complex feature extraction techniques, such as effective connectivity (EC) and phase lag index (PLI). \citet{al2023machine} utilised EC features derived from cortical correlation, which offers insight into the directed interactions between brain regions. \citet{fang2024exploring} employed PLI, which measures phase synchronisation between EEG signals, revealing connectivity patterns across different brain areas. The Discrete Wavelet Transform (DWT) is another widely used feature extraction method. \citet{aldayel2024} and \citet{baghdadi2019dasps,baghdadi2021psychological} applied DWT alongside PSD and other statistical features, such as Hjorth parameters, fractal dimension and spectral entropy. DWT enables the decomposition of EEG signals into multiple frequency bands, making it easier to capture both time and frequency information.

Complexity-based features like Lempel-Ziv complexity and correlation dimension have been explored in specific patient populations. For example, \citet{aderinwale2023two} studied these features identifying lower complexity and correlation in these individuals compared to healthy controls.

Lastly, feature selection techniques such as recursive feature elimination and the wrapper method have been used to optimise machine learning models by selecting the most relevant EEG features. \citet{shikha2021stacked} and \citet{muhammad2022human} employed these methods to enhance classification performance by identifying the most significant features from large feature sets. Overall, the EEG feature extraction techniques in machine learning span a wide range of methods, including time and frequency domain features, complexity measures, connectivity indices and advanced statistical methods 

\begin{table*}
\footnotesize
\centerline{
\begin{threeparttable}
\caption{Summary of Machine Learning Models and Accuracies}
\label{tab:ml_summary}
\begin{tabular}{@{}p{3cm}p{7cm}p{5cm}@{}}
\toprule
\textbf{Study}                   & \textbf{Model}                                                                          & \textbf{Accuracy} \\ \midrule
\citep{shen2022aberrated}             & SVM, RF, ensemble learning                                                               & 97.55\%           \\
\citep{al2021analysis}             & KNN, LDA, NBC, DT, SVM                                                                  & 86.93\%           \\
\citep{arsalan2020electroencephalography}              & Logistic regression, Random Forest, Multilayer Perceptron                               & 78.50\%           \\
\citep{aldayel2024}             & SVM, KNN, LDA, gradient bagging, ADA boost bagging                                      & 87.50\%           \\
\citep{gross2021machine}                & Random Forest                                                                           & 81.25\%           \\
\citep{fang2024exploring}               & XGBoost, CatBoost, LightGBM, and ensemble models                                        & 97.33\%           \\
\citep{al2023machine}              & CNN, LSTM, and CNN + LSTM                                                               & 92.86\%           \\
\citep{aderinwale2023two}           & Support Vector Machine (SVM)                                                            & 68\% PD vs HC     \\
\citep{muhammad2022human}             & MLP, SVM, RF, DT, KNN                                                                   & 94.90\% (9 features), 92.74\% (10 features) \\
\citep{park2021identification}                & SVM, Random Forest, Elastic Net                                                         & 91.03\%           \\
\citep{shikha2021stacked}             & Decision Tree, Random Forest, Stacked Sparse Autoencoder                                & 83.93\% (Stacked Sparse Autoencoder), 70.25\% (Decision Tree) \\
\citep{baghdadi2019dasps, baghdadi2021psychological}          & Stacked Sparse AutoEncoder, KNN, SVM                                                    & 83.50\% (Stacked Sparse AutoEncoder), 81.40\% (KNN), 77.40\% (SVM) \\
\citep{daud2023safe, shing2023multistage} & KNN, SVM, and Decision Tree                                       & 89.5\% accuracy, 89.7\% precision \\
\citep{chen2021eeg}              & SVM: RBF + OVO                                                                          & 92\%              \\
\citep{xie2020anxiety}         & BN + CNN2BN + DBNBN + LDAPL + LDA                                                       & Notable accuracy\\ \bottomrule
\end{tabular}
\begin{tablenotes}
\item \textbf{Abbreviations:} SVM (Support Vector Machine), RF (Random Forest), KNN (K-Nearest Neighbors), LDA (Linear Discriminant Analysis), NBC (Naive Bayes Classifier), DT (Decision Tree), ADA (Adaptive Boosting), XGBoost (Extreme Gradient Boosting), CatBoost (Categorical Boosting), LightGBM (Light Gradient Boosting Machine), CNN (Convolutional Neural Network), LSTM (Long Short-Term Memory), MLP (Multilayer Perceptron), Elastic Net (A regularized regression method), BN (Bayesian Network), DBNBN (Deep Bayesian Network), LDAPL (Latent Dirichlet Allocation for Probability Learning), LDA (Linear Discriminant Analysis).
\end{tablenotes}
\end{threeparttable}}
\end{table*}

\subsection{Machine Learning Algorithms}
\citet{shen2022aberrated} leveraged the power of SVM along with RF and ensemble learning to achieve a high accuracy of 97.55\%. Similarly, \citet{park2021identification} utilised SVM to analyse anxiety disorders, reaching an accuracy of 91.03\% using whole band PSD. \citet{chen2021eeg} also employed SVM with an radial basis function (RBF) kernel and one-versus-one (OVO) strategy, achieving 92\% accuracy. \citet{aderinwale2023two} applied SVM specifically for distinguishing between PD and HC, attaining a 68\% accuracy. \citet{daud2023safe}, \citet{shing2023multistage} combined SVM with KNN and decision trees, resulting in a precision of 89.7\% and an accuracy of up to 89.5\%. These studies highlight the versatility and effectiveness of SVM across various applications.

Ensemble learning proved to be a formidable approach in several studies. \citet{shen2022aberrated} and \citet{fang2024exploring} both utilised ensemble learning techniques, with \citet{fang2024exploring} incorporating XGBoost, CatBoost, LightGBM and other models to achieve an impressive accuracy of 97.33\%. \citet{aldayel2024} applied SVM, KNN, LDA, gradient bagging and ADA boost bagging, achieving 87.50\% accuracy. \citet{gross2021machine} focused on RF, reporting an accuracy of 81.25\%, while \citet{muhammad2022human} employed MLP, SVM, RF, DT and KNN, attaining 94.90\% accuracy with nine features and 92.74\% with ten features. These findings underscore the potential of ensemble methods to enhance prediction accuracy in complex datasets.

Deep learning approaches also featured prominently in this collection of studies. \citet{al2022complexity} explored CNN, LSTM and a combination of CNN + LSTM, resulting in an accuracy of 92.86\%. \citet{baghdadi2019dasps, baghdadi2021psychological} compared a Stacked Sparse AutoEncoder against KNN and SVM, with the Autoencoder outperforming the others at 83.50\% accuracy. \citet{shikha2021stacked} experimented with decision trees, RF and a stacked sparse Autoencoder, with the latter achieving 83.93\% accuracy. \citet{arsalan2020electroencephalography} utilised logistic regression, RF and MLP, reaching an accuracy of 78.50\%. \citet{xie2020anxiety} developed a complex model involving BN, CNN2BN, DBNBN, LDAPL and LDA, achieving notable accuracy. These studies demonstrate the broad range of techniques used to tackle diverse predictive modelling challenges.

Machine learning studies across \autoref{tab:ml_summary} illustrate the various domains that showcased the effectiveness of different models, with SVM and ensemble learning techniques, such as RF and XGBoost, often achieving high accuracy rates. Deep learning methods like CNN and LSTM were also prominently used, with notable success in predicting anxiety disorders.

\section{Discussions and Future Directions}
This review focuses on several key anxiety disorders, including generalised anxiety disorder (GAD) and panic disorders (PD), exploring the use of electroencephalography (EEG) and error-related negativity (ERN) in prior studies for anxiety detection. Despite the prevalence of research utilising GAD signals to study anxiety, there is a notable scarcity of studies concentrating on panic disorders. Most existing studies employ EEG data from all-over-channel electrode sites, yet anxiety disorders primarily involve activity in the frontal electrode sites. This review identifies a gap in the literature concerning the application of ERN with machine learning models, highlighting an area for further exploration.

Feature extraction is a crucial step in EEG analysis, and the majority of anxiety detection approaches in the literature rely on either time-domain or frequency-domain analysis. Some studies have reported significant results using comprehensive electrode site data, yet the most informative features for anxiety detection are often obtained from the frontal regions of the brain where ERN activity is prominent.

Machine learning models have been extensively used for classifying EEG signals in anxiety disorder studies. Support vector machines (SVM) and random forest (RF) models have demonstrated high accuracy and performance in anxiety detection tasks. However, neural network architectures, particularly long short-term memory (LSTM) networks and recurrent neural networks (RNNs) have shown superior accuracy compared to multi-layer perceptrons (MLP) and convolutional neural networks (CNN) for this purpose. These findings suggest that advanced neural networks may offer improved capabilities in capturing the temporal dynamics of EEG signals associated with anxiety disorders.

Lastly, significant advancements have been made in using EEG and ERN for anxiety detection, future research should focus on addressing the identified gaps, such as the application of ERN with machine learning and the investigation of panic disorders. Further exploration of EEG features specific to anxiety subtypes and the utilisation of cutting-edge machine learning models will enhance the precision and applicability of these diagnostic tools in clinical settings.

\section{Conclusion}
This research reviews an optimistic picture for the future of diagnosing and understanding anxiety disorders through a combination of EEG, ERN analysis and machine learning. The studies explored suggest that individuals with anxiety exhibit distinct patterns in their brain activity, particularly in the realm of error detection. Machine learning models, when trained on this EEG data, have shown promising accuracy in differentiating between healthy individuals and those with anxiety disorders. This technology has the potential to revolutionise the diagnostic landscape, offering a non-invasive and potentially objective method for identifying anxiety.

However, there are crucial areas that require further investigation. The studies reviewed primarily focused on four specific disorders: GAD, OCD, PD and SAD. More research is needed to explore the effectiveness of this approach in diagnosing panic disorder, where existing data is scarce. Additionally, while the reviewed models achieved promising results, refining them further is important to improve accuracy and generalisability across different populations.
Looking forward, large-scale clinical trials with diverse participants are essential to validate these findings and establish EEG-machine learning as a reliable diagnostic tool. Further research should also delve deeper into the underlying neural mechanisms that differentiate healthy brains from those with anxiety disorders. This knowledge could pave the way for the development of more targeted treatment approaches. By harnessing the power of EEG and machine learning, researchers hold the potential to significantly improve the lives of millions struggling with anxiety disorders.

\bibliographystyle{abbrvnat}
\bibliography{refere}

\begin{thebibliography}{76}
\providecommand{\natexlab}[1]{#1}
\providecommand{\url}[1]{\texttt{#1}}
\expandafter\ifx\csname urlstyle\endcsname\relax
  \providecommand{\doi}[1]{doi: #1}\else
  \providecommand{\doi}{doi: \begingroup \urlstyle{rm}\Url}\fi

\bibitem[Abd-Alrazaq et~al.(2023)Abd-Alrazaq, AlSaad, Aziz, Ahmed, Denecke, Househ, and Sheikh]{abd2023}
A.~Abd-Alrazaq, R.~AlSaad, S.~Aziz, A.~Ahmed, K.~Denecke, M.~Househ, and J.~Sheikh.
\newblock Correction: Wearable artificial intelligence for anxiety and depression: Scoping review.
\newblock \emph{Journal of Medical Internet Research}, 25:\penalty0 e46233--e4623, 2023.

\bibitem[Aderinwale et~al.(2023)Aderinwale, Tolossa, Kim, Jang, Lee, Jeon, Kim, Yu, and Jeong]{aderinwale2023two}
A.~Aderinwale, G.~B. Tolossa, A.~Y. Kim, E.~H. Jang, Y.-i. Lee, H.~J. Jeon, H.~Kim, H.~Y. Yu, and J.~Jeong.
\newblock Two-channel eeg based diagnosis of panic disorder and major depressive disorder using machine learning and non-linear dynamical methods.
\newblock \emph{Psychiatry Research: Neuroimaging}, 332:\penalty0 111641, 2023.

\bibitem[Al-Ezzi et~al.(2020)Al-Ezzi, Kamel, Faye, and Gunaseli]{al2020review}
A.~Al-Ezzi, N.~Kamel, I.~Faye, and E.~Gunaseli.
\newblock Review of eeg, erp, and brain connectivity estimators as predictive biomarkers of social anxiety disorder.
\newblock \emph{Frontiers in psychology}, 11:\penalty0 730, 2020.

\bibitem[Al-Ezzi et~al.(2021)Al-Ezzi, Kamel, Faye, and Gunaseli]{al2021analysis}
A.~Al-Ezzi, N.~Kamel, I.~Faye, and E.~Gunaseli.
\newblock Analysis of default mode network in social anxiety disorder: Eeg resting-state effective connectivity study.
\newblock \emph{Sensors}, 21\penalty0 (12):\penalty0 4098, 2021.

\bibitem[Al-Ezzi et~al.(2022)Al-Ezzi, Al-Shargabi, Al-Shargie, and Zahary]{al2022complexity}
A.~Al-Ezzi, A.~A. Al-Shargabi, F.~Al-Shargie, and A.~T. Zahary.
\newblock Complexity analysis of eeg in patients with social anxiety disorder using fuzzy entropy and machine learning techniques.
\newblock \emph{IEEE Access}, 10:\penalty0 39926--39938, 2022.

\bibitem[Al-Ezzi et~al.(2023)Al-Ezzi, Kamel, Al-Shargabi, Al-Shargie, Al-Shargabi, Yahya, and Al-Hiyali]{al2023machine}
A.~Al-Ezzi, N.~Kamel, A.~A. Al-Shargabi, F.~Al-Shargie, A.~Al-Shargabi, N.~Yahya, and M.~I. Al-Hiyali.
\newblock Machine learning for the detection of social anxiety disorder using effective connectivity and graph theory measures.
\newblock \emph{Frontiers in psychiatry}, 14:\penalty0 1155812, 2023.

\bibitem[Aldayel and Al-Nafjan(2024)]{aldayel2024}
M.~Aldayel and A.~Al-Nafjan.
\newblock A comprehensive exploration of machine learning techniques for eeg-based anxiety detection.
\newblock \emph{PeerJ Computer Science}, 10:\penalty0 e1829, 2024.
\newblock \doi{10.7717/peerj-cs.1829}.

\bibitem[Arsalan et~al.(2020)Arsalan, Majid, and Anwar]{arsalan2020electroencephalography}
A.~Arsalan, M.~Majid, and S.~M. Anwar.
\newblock Electroencephalography based machine learning framework for anxiety classification.
\newblock In \emph{Intelligent Technologies and Applications: Second International Conference, INTAP 2019, Bahawalpur, Pakistan, November 6--8, 2019, Revised Selected Papers 2}, pages 187--197. Springer, 2020.

\bibitem[{Australian Bureau of Statistics}()]{abs}
{Australian Bureau of Statistics}.
\newblock Mental health.
\newblock URL \url{https://www.abs.gov.au/statistics/health/mental-health}.

\bibitem[Baghdadi et~al.(2019)Baghdadi, Aribi, Fourati, Halouani, Siarry, and Alimi]{baghdadi2019dasps}
A.~Baghdadi, Y.~Aribi, R.~Fourati, N.~Halouani, P.~Siarry, and A.~M. Alimi.
\newblock Dasps: a database for anxious states based on a psychological stimulation.
\newblock \emph{arXiv preprint arXiv:1901.02942}, 2019.

\bibitem[Baghdadi et~al.(2021)Baghdadi, Aribi, Fourati, Halouani, Siarry, and Alimi]{baghdadi2021psychological}
A.~Baghdadi, Y.~Aribi, R.~Fourati, N.~Halouani, P.~Siarry, and A.~Alimi.
\newblock Psychological stimulation for anxious states detection based on eeg-related features.
\newblock \emph{Journal of Ambient Intelligence and Humanized Computing}, 12:\penalty0 8519--8533, 2021.

\bibitem[Banica et~al.(2019)Banica, Sandre, and Weinberg]{banica2019overprotective}
I.~Banica, A.~Sandre, and A.~Weinberg.
\newblock Overprotective/authoritarian maternal parenting is associated with an enhanced error-related negativity (ern) in emerging adult females.
\newblock \emph{International Journal of Psychophysiology}, 137:\penalty0 12--20, 2019.

\bibitem[Barker et~al.(2015)Barker, Troller-Renfree, Pine, and Fox]{barker2015individual}
T.~V. Barker, S.~Troller-Renfree, D.~S. Pine, and N.~A. Fox.
\newblock Individual differences in social anxiety affect the salience of errors in social contexts.
\newblock \emph{Cognitive, Affective, \& Behavioral Neuroscience}, 15:\penalty0 723--735, 2015.

\bibitem[Br{\'a}zdil et~al.(2005)Br{\'a}zdil, Roman, Daniel, and Rektor]{brazdil2005intracerebral}
M.~Br{\'a}zdil, R.~Roman, P.~Daniel, and I.~Rektor.
\newblock Intracerebral error-related negativity in a simple go/nogo task.
\newblock \emph{Journal of Psychophysiology}, 19\penalty0 (4):\penalty0 244--255, 2005.

\bibitem[Carrasco(2012)]{carrasco2012cognitive}
M.~Carrasco.
\newblock \emph{Cognitive Role of Medial PFC in Error Processing: Lessons Learned from Healthy Children and Pediatric OCD, Anxiety, and ASD.}
\newblock PhD thesis, 2012.

\bibitem[Carrasco et~al.(2013)Carrasco, Harbin, Nienhuis, Fitzgerald, Gehring, and Hanna]{carrasco2013increased}
M.~Carrasco, S.~M. Harbin, J.~K. Nienhuis, K.~D. Fitzgerald, W.~J. Gehring, and G.~L. Hanna.
\newblock Increased error-related brain activity in youth with obsessive-compulsive disorder and unaffected siblings.
\newblock \emph{Depression and anxiety}, 30\penalty0 (1):\penalty0 39--46, 2013.

\bibitem[Chen et~al.(2021)Chen, Yu, Belkacem, Lu, Li, Zhang, Wang, Tan, Gao, Shin, et~al.]{chen2021eeg}
C.~Chen, X.~Yu, A.~N. Belkacem, L.~Lu, P.~Li, Z.~Zhang, X.~Wang, W.~Tan, Q.~Gao, D.~Shin, et~al.
\newblock Eeg-based anxious states classification using affective bci-based closed neurofeedback system.
\newblock \emph{Journal of medical and biological engineering}, 41:\penalty0 155--164, 2021.

\bibitem[Cole et~al.(2023)Cole, Mehra, Cibrian, Cummings, Nelson, Hajcak, and Meyer]{cole2023relational}
S.~L. Cole, L.~M. Mehra, E.~Cibrian, E.~M. Cummings, B.~D. Nelson, G.~Hajcak, and A.~Meyer.
\newblock Relational victimization prospectively predicts increases in error-related brain activity and social anxiety in children and adolescents across two years.
\newblock \emph{Developmental Cognitive Neuroscience}, 61:\penalty0 101252, 2023.

\bibitem[Crawford et~al.(2020)Crawford, Moss, Groves, Dowlen, Nelson, Reid, and Oliver]{crawford2020behavioural}
H.~Crawford, J.~Moss, L.~Groves, R.~Dowlen, L.~Nelson, D.~Reid, and C.~Oliver.
\newblock A behavioural assessment of social anxiety and social motivation in fragile x, cornelia de lange and rubinstein-taybi syndromes.
\newblock \emph{Journal of Autism and Developmental Disorders}, 50:\penalty0 127--144, 2020.

\bibitem[Daud et~al.(2023)Daud, Sudirman, and Shing]{daud2023safe}
S.~N. S.~S. Daud, R.~Sudirman, and T.~W. Shing.
\newblock Safe-level smote method for handling the class imbalanced problem in electroencephalography dataset of adult anxious state.
\newblock \emph{Biomedical Signal Processing and Control}, 83:\penalty0 104649, 2023.

\bibitem[de~Bardeci et~al.(2021)de~Bardeci, Ip, and Olbrich]{de2021deep}
M.~de~Bardeci, C.~T. Ip, and S.~Olbrich.
\newblock Deep learning applied to electroencephalogram data in mental disorders: A systematic review.
\newblock \emph{Biological Psychology}, 162:\penalty0 108117, 2021.

\bibitem[Dehaene et~al.(1994)Dehaene, Posner, and Tucker]{dehaene1994localization}
S.~Dehaene, M.~I. Posner, and D.~M. Tucker.
\newblock Localization of a neural system for error detection and compensation.
\newblock \emph{Psychological science}, 5\penalty0 (5):\penalty0 303--305, 1994.

\bibitem[Endrass et~al.(2010)Endrass, Schuermann, Kaufmann, Spielberg, Kniesche, and Kathmann]{endrass2010performance}
T.~Endrass, B.~Schuermann, C.~Kaufmann, R.~Spielberg, R.~Kniesche, and N.~Kathmann.
\newblock Performance monitoring and error significance in patients with obsessive-compulsive disorder.
\newblock \emph{Biological psychology}, 84\penalty0 (2):\penalty0 257--263, 2010.

\bibitem[Endrass et~al.(2014)Endrass, Riesel, Kathmann, and Buhlmann]{endrass2014performance}
T.~Endrass, A.~Riesel, N.~Kathmann, and U.~Buhlmann.
\newblock Performance monitoring in obsessive--compulsive disorder and social anxiety disorder.
\newblock \emph{Journal of abnormal psychology}, 123\penalty0 (4):\penalty0 705, 2014.

\bibitem[Falkenstein et~al.(1991)Falkenstein, Hohnsbein, Hoormann, and Blanke]{falkenstein1991effects}
M.~Falkenstein, J.~Hohnsbein, J.~Hoormann, and L.~Blanke.
\newblock Effects of crossmodal divided attention on late erp components. ii. error processing in choice reaction tasks.
\newblock \emph{Electroencephalography and clinical neurophysiology}, 78\penalty0 (6):\penalty0 447--455, 1991.

\bibitem[Fang et~al.(2024)Fang, Li, Xu, Liu, Chen, Zhu, Luo, Luo, and Zhou]{fang2024exploring}
J.~Fang, G.~Li, W.~Xu, W.~Liu, G.~Chen, Y.~Zhu, Y.~Luo, X.~Luo, and B.~Zhou.
\newblock Exploring abnormal brain functional connectivity in healthy adults, depressive disorder, and generalized anxiety disorder through eeg signals: A machine learning approach for triple classification.
\newblock \emph{Brain Sciences}, 14\penalty0 (3):\penalty0 245, 2024.

\bibitem[Gehring et~al.(1993)Gehring, Goss, Coles, Meyer, and Donchin]{gehring1993neural}
W.~J. Gehring, B.~Goss, M.~G. Coles, D.~E. Meyer, and E.~Donchin.
\newblock A neural system for error detection and compensation.
\newblock \emph{Psychological science}, 4\penalty0 (6):\penalty0 385--390, 1993.

\bibitem[Ghosh and Anwar(2021)]{ghosh2021depression}
S.~Ghosh and T.~Anwar.
\newblock Depression intensity estimation via social media: A deep learning approach.
\newblock \emph{IEEE Transactions on Computational Social Systems}, 8\penalty0 (6):\penalty0 1465--1474, 2021.

\bibitem[Gross et~al.(2021)Gross, Mesgun, Frick, Baumgartl, and Buettner]{gross2021machine}
J.~Gross, F.~Mesgun, J.~Frick, H.~Baumgartl, and R.~Buettner.
\newblock Machine learning-based detection of high trait anxiety using frontal asymmetry characteristics in resting-state eeg recordings.
\newblock \emph{Machine Learning}, 7:\penalty0 12--2021, 2021.

\bibitem[Hajcak et~al.(2003)Hajcak, McDonald, and Simons]{hajcak2003anxiety}
G.~Hajcak, N.~McDonald, and R.~F. Simons.
\newblock Anxiety and error-related brain activity.
\newblock \emph{Biological psychology}, 64\penalty0 (1-2):\penalty0 77--90, 2003.

\bibitem[Hajcak et~al.(2004)Hajcak, McDonald, and Simons]{hajcak2004error}
G.~Hajcak, N.~McDonald, and R.~F. Simons.
\newblock Error-related psychophysiology and negative affect.
\newblock \emph{Brain and cognition}, 56\penalty0 (2):\penalty0 189--197, 2004.

\bibitem[Hanna et~al.(2020)Hanna, Liu, Rough, Surapaneni, Hanna, Arnold, and Gehring]{hanna2020diagnostic}
G.~L. Hanna, Y.~Liu, H.~E. Rough, M.~Surapaneni, B.~S. Hanna, P.~D. Arnold, and W.~J. Gehring.
\newblock A diagnostic biomarker for pediatric generalized anxiety disorder using the error-related negativity.
\newblock \emph{Child Psychiatry \& Human Development}, 51:\penalty0 827--838, 2020.

\bibitem[Hum et~al.(2013)Hum, Manassis, and Lewis]{hum2013neural}
K.~M. Hum, K.~Manassis, and M.~D. Lewis.
\newblock Neural mechanisms of emotion regulation in childhood anxiety.
\newblock \emph{Journal of Child Psychology and Psychiatry}, 54\penalty0 (5):\penalty0 552--564, 2013.

\bibitem[Kaczkurkin(2013)]{kaczkurkin2013effect}
A.~N. Kaczkurkin.
\newblock The effect of manipulating task difficulty on error-related negativity in individuals with obsessive-compulsive symptoms.
\newblock \emph{Biological Psychology}, 93\penalty0 (1):\penalty0 122--131, 2013.

\bibitem[Kujawa et~al.(2016)Kujawa, Weinberg, Bunford, Fitzgerald, Hanna, Monk, Kennedy, Klumpp, Hajcak, and Phan]{kujawa2016error}
A.~Kujawa, A.~Weinberg, N.~Bunford, K.~D. Fitzgerald, G.~L. Hanna, C.~S. Monk, A.~E. Kennedy, H.~Klumpp, G.~Hajcak, and K.~L. Phan.
\newblock Error-related brain activity in youth and young adults before and after treatment for generalized or social anxiety disorder.
\newblock \emph{Progress in Neuro-Psychopharmacology and Biological Psychiatry}, 71:\penalty0 162--168, 2016.

\bibitem[Larson et~al.(2013)Larson, Clawson, Clayson, and Baldwin]{larson2013cognitive}
M.~J. Larson, A.~Clawson, P.~E. Clayson, and S.~A. Baldwin.
\newblock Cognitive conflict adaptation in generalized anxiety disorder.
\newblock \emph{Biological Psychology}, 94\penalty0 (2):\penalty0 408--418, 2013.

\bibitem[Li et~al.(2020)Li, Ghosh, Joshi, and Oviatt]{li2020lstm}
Y.~Li, S.~Ghosh, J.~Joshi, and S.~Oviatt.
\newblock Lstm-dnn based approach for pain intensity and protective behaviour prediction.
\newblock In \emph{2020 15th IEEE International Conference on Automatic Face and Gesture Recognition (FG 2020)}, pages 819--823. IEEE, 2020.

\bibitem[Li et~al.(2021)Li, Ghosh, and Joshi]{li2021plaan}
Y.~Li, S.~Ghosh, and J.~Joshi.
\newblock Plaan: pain level assessment with anomaly-detection based network.
\newblock \emph{Journal on Multimodal User Interfaces}, 15\penalty0 (4):\penalty0 359--372, 2021.

\bibitem[Liu et~al.(2023)Liu, Li, Huang, Jiang, Luo, and Xu]{liu2023enhancing}
W.~Liu, G.~Li, Z.~Huang, W.~Jiang, X.~Luo, and X.~Xu.
\newblock Enhancing generalized anxiety disorder diagnosis precision: Mstcnn model utilizing high-frequency eeg signals.
\newblock \emph{Frontiers in Psychiatry}, 14:\penalty0 1310323, 2023.

\bibitem[Lo et~al.(2017)Lo, Schroder, Fisher, Durbin, Fitzgerald, Danovitch, and Moser]{lo2017associations}
S.~L. Lo, H.~S. Schroder, M.~E. Fisher, C.~E. Durbin, K.~D. Fitzgerald, J.~H. Danovitch, and J.~S. Moser.
\newblock Associations between disorder-specific symptoms of anxiety and error-monitoring brain activity in young children.
\newblock \emph{Journal of Abnormal Child Psychology}, 45:\penalty0 1439--1448, 2017.

\bibitem[Luo et~al.(2024)Luo, Zhou, Fang, Cherif-Riahi, Li, and Shen]{luo2024integrating}
X.~Luo, B.~Zhou, J.~Fang, Y.~Cherif-Riahi, G.~Li, and X.~Shen.
\newblock Integrating eeg and ensemble learning for accurate grading and quantification of generalized anxiety disorder: A novel diagnostic approach.
\newblock \emph{Diagnostics}, 14\penalty0 (11):\penalty0 1122, 2024.

\bibitem[Mazlan et~al.(2024)Mazlan, Sukor, Adom, and Jamaluddin]{mazlan2024review}
M.~R. Mazlan, A.~S.~A. Sukor, A.~H. Adom, and R.~Jamaluddin.
\newblock Review of analysis of eeg signals for stress detection.
\newblock In \emph{AIP Conference Proceedings}, volume 2934. AIP Publishing, 2024.

\bibitem[Meyer(2016)]{meyer2016}
A.~Meyer.
\newblock Developing psychiatric biomarkers: a review focusing on the error-related negativity as a biomarker for anxiety.
\newblock \emph{Current Treatment Options in Psychiatry}, 3\penalty0 (4):\penalty0 356--364, 2016.
\newblock \doi{10.1007/s40501-016-0094-5}.

\bibitem[Meyer(2017)]{meyer2017biomarker}
A.~Meyer.
\newblock A biomarker of anxiety in children and adolescents: A review focusing on the error-related negativity (ern) and anxiety across development.
\newblock \emph{Developmental cognitive neuroscience}, 27:\penalty0 58--68, 2017.

\bibitem[Meyer et~al.(2012)Meyer, Weinberg, Klein, and Hajcak]{meyer2012development}
A.~Meyer, A.~Weinberg, D.~N. Klein, and G.~Hajcak.
\newblock The development of the error-related negativity (ern) and its relationship with anxiety: Evidence from 8 to 13 year-olds.
\newblock \emph{Developmental Cognitive Neuroscience}, 2\penalty0 (1):\penalty0 152--161, 2012.

\bibitem[Meyer et~al.(2015)Meyer, Hajcak, Torpey-Newman, Kujawa, and Klein]{meyer2015enhanced}
A.~Meyer, G.~Hajcak, D.~C. Torpey-Newman, A.~Kujawa, and D.~N. Klein.
\newblock Enhanced error-related brain activity in children predicts the onset of anxiety disorders between the ages of 6 and 9.
\newblock \emph{Journal of abnormal psychology}, 124\penalty0 (2):\penalty0 266, 2015.

\bibitem[Michael et~al.(2021)Michael, Wang, Kaur, Fitzgerald, Fitzgibbon, and Hoy]{michael2021eeg}
J.~A. Michael, M.~Wang, M.~Kaur, P.~B. Fitzgerald, B.~M. Fitzgibbon, and K.~E. Hoy.
\newblock Eeg correlates of attentional control in anxiety disorders: A systematic review of error-related negativity and correct-response negativity findings.
\newblock \emph{Journal of affective disorders}, 291:\penalty0 140--153, 2021.

\bibitem[Minkowski et~al.(2021)Minkowski, Mai, and Gurve]{minkowski2021feature}
L.~Minkowski, K.~V. Mai, and D.~Gurve.
\newblock Feature extraction to identify depression and anxiety based on eeg.
\newblock In \emph{2021 43rd Annual International Conference of the IEEE Engineering in Medicine \& Biology Society (EMBC)}, pages 6322--6325. IEEE, 2021.

\bibitem[Mohan and Perumal(2023)]{mohan2023classification}
R.~Mohan and S.~Perumal.
\newblock Classification and detection of cognitive disorders like depression and anxiety utilizing deep convolutional neural network (cnn) centered on eeg signal.
\newblock \emph{Traitement Du Signal}, 40\penalty0 (3), 2023.

\bibitem[Moser et~al.(2013)Moser, Moran, Schroder, Donnellan, and Yeung]{moser2013relationship}
J.~S. Moser, T.~P. Moran, H.~S. Schroder, M.~B. Donnellan, and N.~Yeung.
\newblock On the relationship between anxiety and error monitoring: a meta-analysis and conceptual framework.
\newblock \emph{Frontiers in human neuroscience}, 7:\penalty0 466, 2013.

\bibitem[Mou et~al.(2024)Mou, Yan, Shen, Shuai, Li, Shen, and Shen]{mou2024prolonged}
S.~Mou, S.~Yan, S.~Shen, Y.~Shuai, G.~Li, Z.~Shen, and P.~Shen.
\newblock Prolonged disease course leads to impaired brain function in anxiety disorder: a resting state eeg study.
\newblock \emph{Neuropsychiatric Disease and Treatment}, pages 1409--1419, 2024.

\bibitem[Mughal et~al.(2020)Mughal, Devadas, Ardman, Levis, Go, and Gaynes]{mughal2020systematic}
A.~Y. Mughal, J.~Devadas, E.~Ardman, B.~Levis, V.~F. Go, and B.~N. Gaynes.
\newblock A systematic review of validated screening tools for anxiety disorders and ptsd in low to middle income countries.
\newblock \emph{BMC psychiatry}, 20:\penalty0 1--18, 2020.

\bibitem[Muhammad and Al-Ahmadi(2022)]{muhammad2022human}
F.~Muhammad and S.~Al-Ahmadi.
\newblock Human state anxiety classification framework using eeg signals in response to exposure therapy.
\newblock \emph{Plos one}, 17\penalty0 (3):\penalty0 e0265679, 2022.

\bibitem[{National Institute of Mental Health}()]{nimh}
{National Institute of Mental Health}.
\newblock Depression.
\newblock URL \url{https://www.nimh.nih.gov/health/topics/depression}.

\bibitem[Page et~al.(2021)Page, McKenzie, Bossuyt, Boutron, Hoffmann, Mulrow, Shamseer, Tetzlaff, Akl, Brennan, et~al.]{page2021prisma}
M.~J. Page, J.~E. McKenzie, P.~M. Bossuyt, I.~Boutron, T.~C. Hoffmann, C.~D. Mulrow, L.~Shamseer, J.~M. Tetzlaff, E.~A. Akl, S.~E. Brennan, et~al.
\newblock The prisma 2020 statement: an updated guideline for reporting systematic reviews.
\newblock \emph{BMJ}, 372:\penalty0 n71, 2021.

\bibitem[Park et~al.(2021)Park, Jeong, Oh, Choi, Jung, Lee, Lee, and Choi]{park2021identification}
S.~M. Park, B.~Jeong, D.~Y. Oh, C.-H. Choi, H.~Y. Jung, J.-Y. Lee, D.~Lee, and J.-S. Choi.
\newblock Identification of major psychiatric disorders from resting-state electroencephalography using a machine learning approach.
\newblock \emph{Frontiers in Psychiatry}, 12:\penalty0 707581, 2021.

\bibitem[Qi et~al.(2023)Qi, Xu, and Li]{qi2023neuroimaging}
X.~Qi, W.~Xu, and G.~Li.
\newblock Neuroimaging study of brain functional differences in generalized anxiety disorder and depressive disorder.
\newblock \emph{Brain Sciences}, 13\penalty0 (9):\penalty0 1282, 2023.

\bibitem[Rabinak et~al.(2013)Rabinak, Holman, Angstadt, Kennedy, Hajcak, and Phan]{rabinak2013neural}
C.~A. Rabinak, A.~Holman, M.~Angstadt, A.~E. Kennedy, G.~Hajcak, and K.~L. Phan.
\newblock Neural response to errors in combat-exposed returning veterans with and without post-traumatic stress disorder: A preliminary event-related potential study.
\newblock \emph{Psychiatry Research: Neuroimaging}, 213\penalty0 (1):\penalty0 71--78, 2013.

\bibitem[Riesel et~al.(2014)Riesel, Kathmann, and Endrass]{riesel2014overactive}
A.~Riesel, N.~Kathmann, and T.~Endrass.
\newblock Overactive performance monitoring in obsessive--compulsive disorder is independent of symptom expression.
\newblock \emph{European archives of psychiatry and clinical neuroscience}, 264:\penalty0 707--717, 2014.

\bibitem[Riesel et~al.(2019)Riesel, Klawohn, Gr{\"u}tzmann, Kaufmann, Heinzel, Bey, Lennertz, Wagner, and Kathmann]{riesel2019error}
A.~Riesel, J.~Klawohn, R.~Gr{\"u}tzmann, C.~Kaufmann, S.~Heinzel, K.~Bey, L.~Lennertz, M.~Wagner, and N.~Kathmann.
\newblock Error-related brain activity as a transdiagnostic endophenotype for obsessive-compulsive disorder, anxiety and substance use disorder.
\newblock \emph{Psychological medicine}, 49\penalty0 (7):\penalty0 1207--1217, 2019.

\bibitem[Roh et~al.(2017)Roh, Chang, Yoo, Shin, and Kim]{roh2017modulation}
D.~Roh, J.-G. Chang, S.~Yoo, J.~Shin, and C.-H. Kim.
\newblock Modulation of error monitoring in obsessive--compulsive disorder by individually tailored symptom provocation.
\newblock \emph{Psychological medicine}, 47\penalty0 (12):\penalty0 2071--2080, 2017.

\bibitem[Sanei and Chambers(2007)]{sanei2007introduction}
S.~Sanei and J.~Chambers.
\newblock Introduction to eeg.
\newblock \emph{EEG signal processing}, pages 1--34, 2007.

\bibitem[Schacter et~al.(2011)Schacter, Gilbert, Wegner, and Hood]{schacter2011psychology}
D.~Schacter, D.~Gilbert, D.~Wegner, and B.~M. Hood.
\newblock \emph{Psychology: European Edition}.
\newblock Macmillan International Higher Education, 2011.

\bibitem[Shen et~al.(2022)Shen, Li, Fang, Zhong, Wang, Sun, and Shen]{shen2022aberrated}
Z.~Shen, G.~Li, J.~Fang, H.~Zhong, J.~Wang, Y.~Sun, and X.~Shen.
\newblock Aberrated multidimensional eeg characteristics in patients with generalized anxiety disorder: a machine-learning based analysis framework.
\newblock \emph{Sensors}, 22\penalty0 (14):\penalty0 5420, 2022.

\bibitem[Shikha et~al.(2021)Shikha, Agrawal, Anwar, and Sethia]{shikha2021stacked}
Shikha, M.~Agrawal, M.~A. Anwar, and D.~Sethia.
\newblock Stacked sparse autoencoder and machine learning based anxiety classification using eeg signals.
\newblock In \emph{Proceedings of the First International Conference on AI-ML Systems}, pages 1--7, 2021.

\bibitem[Shing et~al.(2023)Shing, Sudirman, Daud, Razak, Zakaria, and Mahmood]{shing2023multistage}
T.~W. Shing, R.~Sudirman, S.~N. S.~S. Daud, M.~A.~A. Razak, N.~A. Zakaria, and N.~H. Mahmood.
\newblock Multistage anxiety state recognition based on eeg signal using safe-level smote.
\newblock In \emph{Journal of Physics: Conference Series}, volume 2622, page 012010. IOP Publishing, 2023.

\bibitem[Torpey et~al.(2013)Torpey, Hajcak, Kim, Kujawa, Dyson, Olino, and Klein]{torpey2013error}
D.~C. Torpey, G.~Hajcak, J.~Kim, A.~J. Kujawa, M.~W. Dyson, T.~M. Olino, and D.~N. Klein.
\newblock Error-related brain activity in young children: Associations with parental anxiety and child temperamental negative emotionality.
\newblock \emph{Journal of Child Psychology and Psychiatry}, 54\penalty0 (8):\penalty0 854--862, 2013.

\bibitem[Wang et~al.(2022)Wang, Fang, Xu, Zhong, Li, Li, and Li]{wang2022difference}
J.~Wang, J.~Fang, Y.~Xu, H.~Zhong, J.~Li, H.~Li, and G.~Li.
\newblock Difference analysis of multidimensional electroencephalogram characteristics between young and old patients with generalized anxiety disorder.
\newblock \emph{Frontiers in Human Neuroscience}, 16:\penalty0 1074587, 2022.

\bibitem[Weinberg et~al.(2012)Weinberg, Klein, and Hajcak]{weinberg2012increased}
A.~Weinberg, D.~N. Klein, and G.~Hajcak.
\newblock Increased error-related brain activity distinguishes generalized anxiety disorder with and without comorbid major depressive disorder.
\newblock \emph{Journal of abnormal psychology}, 121\penalty0 (4):\penalty0 885, 2012.

\bibitem[Weinberg et~al.(2016)Weinberg, Meyer, Hale-Rude, Perlman, Kotov, Klein, and Hajcak]{weinberg2016error}
A.~Weinberg, A.~Meyer, E.~Hale-Rude, G.~Perlman, R.~Kotov, D.~N. Klein, and G.~Hajcak.
\newblock Error-related negativity (ern) and sustained threat: Conceptual framework and empirical evaluation in an adolescent sample.
\newblock \emph{Psychophysiology}, 53\penalty0 (3):\penalty0 372--385, 2016.

\bibitem[Wiswede et~al.(2009)Wiswede, M{\"u}nte, Goschke, and R{\"u}sseler]{wiswede2009modulation}
D.~Wiswede, T.~F. M{\"u}nte, T.~Goschke, and J.~R{\"u}sseler.
\newblock Modulation of the error-related negativity by induction of short-term negative affect.
\newblock \emph{Neuropsychologia}, 47\penalty0 (1):\penalty0 83--90, 2009.

\bibitem[{World Health Organization}(2017)]{who2017}
{World Health Organization}.
\newblock \emph{Depression and Other Common Mental Disorders: Global Health Estimates}.
\newblock World Health Organization, Geneva, Switzerland, 2017.

\bibitem[Xiao et~al.(2011)Xiao, Wang, Zhang, Li, Tang, Wang, Fan, and Fromson]{xiao2011error}
Z.~Xiao, J.~Wang, M.~Zhang, H.~Li, Y.~Tang, Y.~Wang, Q.~Fan, and J.~A. Fromson.
\newblock Error-related negativity abnormalities in generalized anxiety disorder and obsessive--compulsive disorder.
\newblock \emph{Progress in Neuro-Psychopharmacology and Biological Psychiatry}, 35\penalty0 (1):\penalty0 265--272, 2011.

\bibitem[Xie et~al.(2020)Xie, Yang, Lu, Zheng, Fan, Bi, Li, et~al.]{xie2020anxiety}
Y.~Xie, B.~Yang, X.~Lu, M.~Zheng, C.~Fan, X.~Bi, Y.~Li, et~al.
\newblock Anxiety and depression diagnosis method based on brain networks and convolutional neural networks.
\newblock In \emph{2020 42nd Annual International Conference of the IEEE Engineering in Medicine \& Biology Society (EMBC)}, pages 1503--1506. IEEE, 2020.

\bibitem[Yadawad et~al.(2024)Yadawad, Pandey, Mallibhat, and Mudenagudi]{yadawad2024predicting}
P.~R. Yadawad, L.~Pandey, K.~Mallibhat, and U.~Mudenagudi.
\newblock Predicting anxiety among young adults using machine learning algorithms.
\newblock In \emph{2024 IEEE 9th International Conference for Convergence in Technology (I2CT)}, pages 1--8. IEEE, 2024.

\bibitem[Zambrano-Vazquez and Allen(2014)]{zambrano2014differential}
L.~Zambrano-Vazquez and J.~J. Allen.
\newblock Differential contributions of worry, anxiety, and obsessive compulsive symptoms to ern amplitudes in response monitoring and reinforcement learning tasks.
\newblock \emph{Neuropsychologia}, 61:\penalty0 197--209, 2014.

\end{thebibliography}

\end{document}